\documentclass[letterpaper,10pt,twocolumn]{article}
\usepackage{graphicx}
\usepackage[hyphens]{url}
\usepackage[hidelinks]{hyperref}
\hypersetup{breaklinks=true}
\usepackage{cite}
\usepackage{makecell}
\usepackage{multirow}
\usepackage{xcolor}
\usepackage{tabularx}
\usepackage{booktabs}
\graphicspath{ {images/} }

\title{Technical Report: A Toolkit for Runtime Detection of Userspace Implants}
\author{
  {\rm J. Aaron Pendergrass}${}^\dag$ \\
  \and
  {\rm Nathan Hull}${}^\dag$ \\
  \and
  {\rm John Clemens}${}^\dag$ \\
  \and
  {\rm Sarah Helble}${}^\dag$ \\
  \and
  {\rm Mark Thober}${}^\dag$ \\
  \and
  {\rm Kathleen McGill}${}^\dag$ \\
  \and
  {\rm Machon Gregory}${}^\ddag$ \\
  \and
  {\rm Peter Loscocco}${}^\ddag$ \\
  \and
  ${}^\dag$ Johns Hopkins Applied Physics Laboratory
  \and
  ${}^\ddag$ National Security Agency
}

\begin{document}
\maketitle
\begin{abstract}
     This paper presents the Userspace Integrity Measurement Toolkit (USIM
Toolkit), a set of integrity measurement collection tools capable of
detecting advanced malware threats, such as memory-only implants, that
evade many traditional detection tools. Userspace integrity
measurement validates that a platform is free from subversion by
validating that the current state of the platform is consistent with a
set of invariants. The invariants enforced by the USIM Toolkit are
carefully chosen based on the expected behavior of userspace, and key
behaviors of advanced malware.  Userspace integrity measurement may be
combined with existing filesystem and kernel integrity measurement
approaches to provide stronger guarantees that a platform is executing
the expected software and that the software is in an expected state.

\end{abstract}

\section{Introduction}
\label{sec:intro}

Modern malware often relies on subtle subversions of the runtime
environment of userspace processes to maintain an adversary's foothold
on victim systems. These \emph{implants} elude most traditional
detection mechanisms while providing a range of features such as
execution of arbitrary programs, bulk data transfers, victim
monitoring, network discovery, persistent storage, proxied
communications, and command and control. In particular,
memory-only implants are able to avoid many popular defensive
technologies based on filesystem scanning. The primary contribution of
this paper is the introduction of userspace integrity measurement, an
approach to filling this gap in malware defense, and the USIM Toolkit,
an initial implementation of userspace integrity measurement with a
demonstrated ability to detect core techniques of advanced implants.
Userspace integrity measurement is based on a principled evaluation of
the userspace abstractions provided by an operating system and the
violations of these abstractions that allow implants to operate
outside the visibility of traditional system monitoring tools.

\subsection{Motivation}
\label{sec:motivation}
The motivation for this work is the conviction that memory-only
implants are active in the wild and often undetected. Memory-only
attacks in general are becoming prevalent
\cite{avt, livingofftheland, crowdstrikecasebook} and go by many names:
advanced volatile threat, fileless, living off the land, malware-free,
memory-only, and non-malware. While detection and reporting of such
attacks is increasing, the most notable report of a memory-only
implant may be Georg Wicherski's talk at SyScan 2014 about Procedure
Linkage Table (PLT) infections and Evanescent Bat\cite{evanescentbat}.

Evanescent Bat was an intrusion into a large European IT company in
mid-2013 in which many Linux servers were compromised. The adversary
injected a memory-only implant into running processes from the
post-exploitation shell that infected the PLT of those processes. The
implant was highly sophisticated: it used \texttt{gcc} plugins to
scrub local variables as they went out of scope and to encrypt global
and heap variables, in addition to handling many different PLT
implementations and corner cases. The intrusion went undetected until
the adversarial operators mistakenly revealed their presence and the
implant.

\subsection{Approach}
Userspace integrity measurement and appraisal is the process of
collecting evidence of the current state of the basic abstractions
provided by an operating system, and evaluating this evidence for
violations of invariants that indicate deviations from expected
runtime behavior. The primary abstractions we consider are
namespaces, filesystems, networking and inter-process communication
channels, environment variables, virtual memory management, and
runtime linker/loaders. Implants take advantage of subtle divergences
between application developers' understandings of these abstractions
and the actual behavior of the operating system. These divergences may
be caused by errors in the operating system implementation,
ambiguities in the specification, or misunderstandings by the
application developers. The premise of userspace measurement is that
these implant behaviors create observable effects in the point-in-time
state of an infected system that are unlikely to be caused by benign
software.

This paper introduces and evaluates the USIM Toolkit as a
proof-of-concept implementation for userspace integrity measurement
targeting the GNU/Linux operating system. Section \ref{sec:userspace}
defines userspace and describes our heuristics for including data as
part of userspace integrity measurement. Section
\ref{sec:adversary-model} describes the adversary model considered
directly by userspace integrity measurement, and how the USIM Toolkit
can be combined with other measurements to protect against more
powerful adversaries. Section \ref{sec:implant-techniques} describes a
range of known implant techniques and how they are likely to impact
userspace. Section \ref{sec:implementation} describes the specific
measurement and appraisal capabilities currently implemented in the
USIM Toolkit. Section \ref{sec:evaluation} evaluates the USIM
Toolkit's ability to detect proof-of-concept implementations
illustrating the techniques of Section \ref{sec:implant-techniques},
and gives initial performance benchmarks for the USIM Toolkit. Section
\ref{sec:related-work} considers our approach in the context of the
large body of existing work combating malware. Section
\ref{sec:future-work} describes the primary limitations and areas for
future work in the current USIM Toolkit implementation. Section
\ref{sec:conclusion} summarizes our approach to userspace measurement
and the primary contributions of this paper.


\section{What Is Userspace}
\label{sec:userspace}
Userspace is everything, aside from the executing kernel, that is
needed for the system to run, persist across reboots, and correctly
perform operations on behalf of the user. A complete userspace integrity
measurement should reflect the security relevant state of all aspects
of a system other than the kernel itself. Rather than attempt to
produce such a measurement, we focus on providing a core set of
measurement tools aimed at bootstrapping trust from a kernel
measurement to the underlying runtime provided for all processes on a
system.

Userspace refers to those aspects of a computer system that are built
using abstractions provided by an operating system's kernel. The exact
features and mechanisms of userspace may vary from system to
system. From an attacker's perspective, userspace provides a rich set
of opportunities to cause the system to diverge from correct operation
without requiring direct modification of the kernel. An expansive
approach to userspace measurement would reveal this variety of
compromises by examining all elements of userspace, recording their
attributes and interrelationships, and appraising these measurements
against expected values for the system. Such a measurement would
reflect
\begin{itemize}
     \item The complete memory state of every executing process,
     including application specific data and common process runtime
     structures
     \item Inter-process communication mechanisms including network
     state, shared memory regions, files opened by processes, and
     pipes
     \item The complete state of the file system including all
     subtrees, executable images, system libraries, data files, and
     program configurations
     \item Configuration data needed for system administration,
     including user accounts and application specific configuration
     semantics such as a webserver's configuration file
     \item The set of configured devices and how they are exposed
     (e.g., via the \texttt{/dev} virtual filesystem)
     \item Policy configurations used for access control, network
     management, boot-time process execution, or other system services
     \item The state of kernel-level data reflecting the current
     configuration of the system such as date and time information,
     process privileges, memory maps, namespaces
\end{itemize}

Such a measurement would be extremely challenging both to collect and
to verify. It would likely be extremely large and depend on
significant application-specific knowledge to reflect the internal
data of the processes that happen to be running on a given
system. Thus, our definition of userspace integrity measurement is
narrower than this. Rather than reflect all aspects of userspace in a
single measurement, we focus on capturing the dynamic state of the
common runtime environment. Which data we include in the userspace
integrity measurement is based on the following heuristics:

\begin{itemize}
     \item \textbf{Include:} Well-formedness conditions that apply to
     all processes executing on a system should be verifiable based on
     userspace measurement.
     \item \textbf{Include:} The current relationships between
     executing processes should be reflected in userspace measurement.
     \item \textbf{Include:} The values of kernel-level data
     structures that directly govern how processes interact with
     system resources should be reflected in userspace measurement.
     \item \textbf{Include:} Cryptographic hashes of critical system
     files that must maintain bit-for-bit equality with a trusted
     baseline for correct system operation should be included in
     userspace measurement. This includes system executables, shared
     libraries, and some configuration files.
     \item \textbf{Exclude:} Well-formedness of kernel-level data
     structures should not be reflected in userspace integrity
     measurement.
     \item \textbf{Exclude:} Application-specific semantics are best
     verified by application-specific measurements and thus should not
     be included in the generic userspace integrity measurement.

\end{itemize}

For example, we include measurements of the PLTs of running processes
as part of userspace integrity measurement, but exclude the content of
data structures within a process's runtime heap. The correctness of the
PLT is core to what it means to be a well-formed process, while the
semantics of the data held inside the process heap is application
specific and excluded from our measurement. Similarly, data maintained
by the kernel that define the privileges of processes, such as the
user id associated with each process, should be included in userspace
measurement, but the details of how these data are stored in the
kernel, e.g., the tree of \texttt{struct task\_struct} instances
maintained by the Linux kernel, should not.

This approach leads to a model for userspace integrity that supports
the detection the kinds of advanced malware threats described in
Section \ref{sec:motivation}, but does not present the technical and
adminsitrative challenges of representing the allowable internal
states of all programs that may be present on a system.


\section{Adversary Model}
\label{sec:adversary-model}
Userspace measurement targets an adversary that is able to arbitrarily
modify the memory of any normal system process but is unable to modify
the USIM Toolkit or its trusted computing base. In particular, the
adversary is unable to corrupt the operating system kernel hosting
both the measurement tools and the victim processes. In existing
systems, most adversaries capable of modifying arbitrary processes are
likely to be able to also modify the measurement tools or operating
system kernel. To protect against realistic adversaries our approach
relies on the concept of nested measurements and isolated execution
environments provided by a measurement and attestation system such as
Maat\cite{maat}.

Conversely, the adversaries considered by the USIM Toolkit could
employ application-specific implants to meet their goals. For example,
an implant could modify function pointers defined by a particular
network service daemon in order to maintain a presence on a
system. Such an implant would not be detected by the USIM Toolkit,
because the USIM Toolkit does not inspect application specific
data. Detection of such an attack would require application-specific
measurement in addition to userspace, kernel, and lower-level
measurements. While we believe there is ample opportunity for
developing measurement strategies capable of verifying
application-specific properties, (a) these measurements would rely on
the guarantees provided by a system like the USIM Toolkit, and (b)
mitigating the more general attack classes should be higher priority
as they tend to provide greater robustness and portability for the
adversary.

The key to generalizing userspace integrity measurement to more
interesting classes of adversaries is the ability of one measurement
to provide evidence that another measurement was correctly
performed. For example, a measurement capability built into the OS
kernel can provide evidence that the USIM Toolkit is correctly
installed and executed. Similarly, a measurement capability such as
the Linux Kernel Integrity Measurer (LKIM) \cite{lkim} may use virtual
machine introspection to provide
evidence that the kernel-level measurement capability is
reliable. Each measurement in the chain supports trust in the
measurements above it. Ultimately, this chain of measurements should
be rooted in low-level cryptographic guarantees such as those provided
by the trusted boot process and trusted platform
module\cite{tboot, tpm}. The formalization of protocols for
specifying appropriate chains of measurements for a given use case is
an area of ongoing research \cite{bundling} which could also be
leveraged by implementations of the USIM Toolkit to guarantee
integrity.


\section{Implant Techniques}
\label{sec:implant-techniques}

All implants have the fundamental goal of providing an adversary
stealthy access to a victim system. Inherent in achieving this goal is
the requirement for execution. An implant that cannot execute cannot
serve its purpose. Implants may satisfy this requirement using
various techniques that we divide into two categories:
simple and sophisticated.

Simple techniques rely on ``hiding in plain sight'' to evade
detection. They use system interfaces in a generally expected way to a
nefarious end and succeed when users and administrators fail to
carefully examine the primary state of the system (e.g., processes
running and files on the file system). Such techniques include
replacing system binaries with malicious binaries, file infections,
and more generally file modifications, and they fall under Joanna
Rutkowska's type 0 malware classification\cite{rutkowska}. These
techniques are uninteresting in this context because they have been
well studied\cite{unix-rootkits} and may be detected by file integrity
checkers (e.g. IMA and Tripwire) or other security and system auditing
tools, including antivirus software.

Sophisticated techniques use system interfaces in unusual ways to
evade detection. These techniques go undetected without deeper
examination of the primary state of the system and fall under
Rutkowska's type I and type II malware
classifications\cite{rutkowska}. It is impossible to anticipate all
conceivable techniques in this category, but we have identified some
common techniques implants may use to achieve execution. We also chose
to examine a relatively new technique for concealment and a privileged
resource acquisition technique that may facilitate communication. The
techniques are
\begin{itemize}
     \item Process Text Segment Modification
     \item Global Offset Table/Procedure Linkage Table Hooking
     \item Shared Object Injection
     \item Thread Injection
     \item Namespace Manipulation
     \item File Descriptor Passing
\end{itemize}

These techniques cover a core set of userspace effects the USIM
Toolkit needs to be able to detect and distinguish from benign system
behavior in order to detect an implant. The first three are common
userland rootkit techniques that abuse existing
processes\cite[Chapter~25]{artofmemoryforensics} to achieve
execution. Thread injection has some overlap with those techniques
(i.e., it may be achieved using any of the previous three), but has a
distinct effect on userspace. Namespace manipulation is a relatively
new means of achieving stealth that may become increasingly relevant
as containers become more common\cite{vmsvscontainers}. File
descriptor passing is a long-standing feature of UNIX domain sockets
that may enable an implant to acquire or transfer privileged resource
access. Our understanding of these techniques directly informed the
prioritization of measurements included in the initial implementation
of the USIM Toolkit.

\subsection{Process Text Segment Modification}
Text segment modification is one of the simplest approaches an implant
can take to maintaining execution within a legitimate process. For
most programs, executable code is mapped directly into the process
memory space from the ``.text'' sections of the program binary and
supporting shared libraries. Notable counter-examples to this rule are
the Procedure Linkage Tables (PLTs) of dynamically linked processes
that are generated by the runtime linker/loader (discussed below in Section
\ref{got-plt-hooking}), and just-in-time compiled code that is
generated by many interpreters in dynamically allocated heap buffers.
Modifying executable code in place gives the implant all the
permissions of the host process, allows the implant to intercept
communications intended for the host process, and evades some basic
detection approaches by not creating new executable memory regions.

By default most programs' load segments are marked as
executable/non-writeable. This causes the runtime loader to request
that the operating system map these regions without the writable bit
set in the corresponding page table entries, which will prevent an
implant from na\"{\i}vely attempting to overwrite a process's code. Many
implants overcome this limitation by employing a code-reuse attack,
such as Return-oriented Programming (ROP)\cite{rop}, to remap part of
the text segment then inject and jump to the implant payload. For
consistency, the payload should then remap the region as non-writable
before entering its main program (although not all implants are this
careful).

This approach is observable, even after the fact, because it modifies
memory pages that should be identical to the on-disk representation in
the program binary. Given the on-disk binaries of the program and all of
its shared object dependencies, a simple comparison with process memory
reveals any modifications\cite[Chapter~25]{artofmemoryforensics}.

\subsection{GOT/PLT Hooking}
\label{got-plt-hooking}
\begin{figure}[h]
     \begin{center}
          \includegraphics[width=7cm]{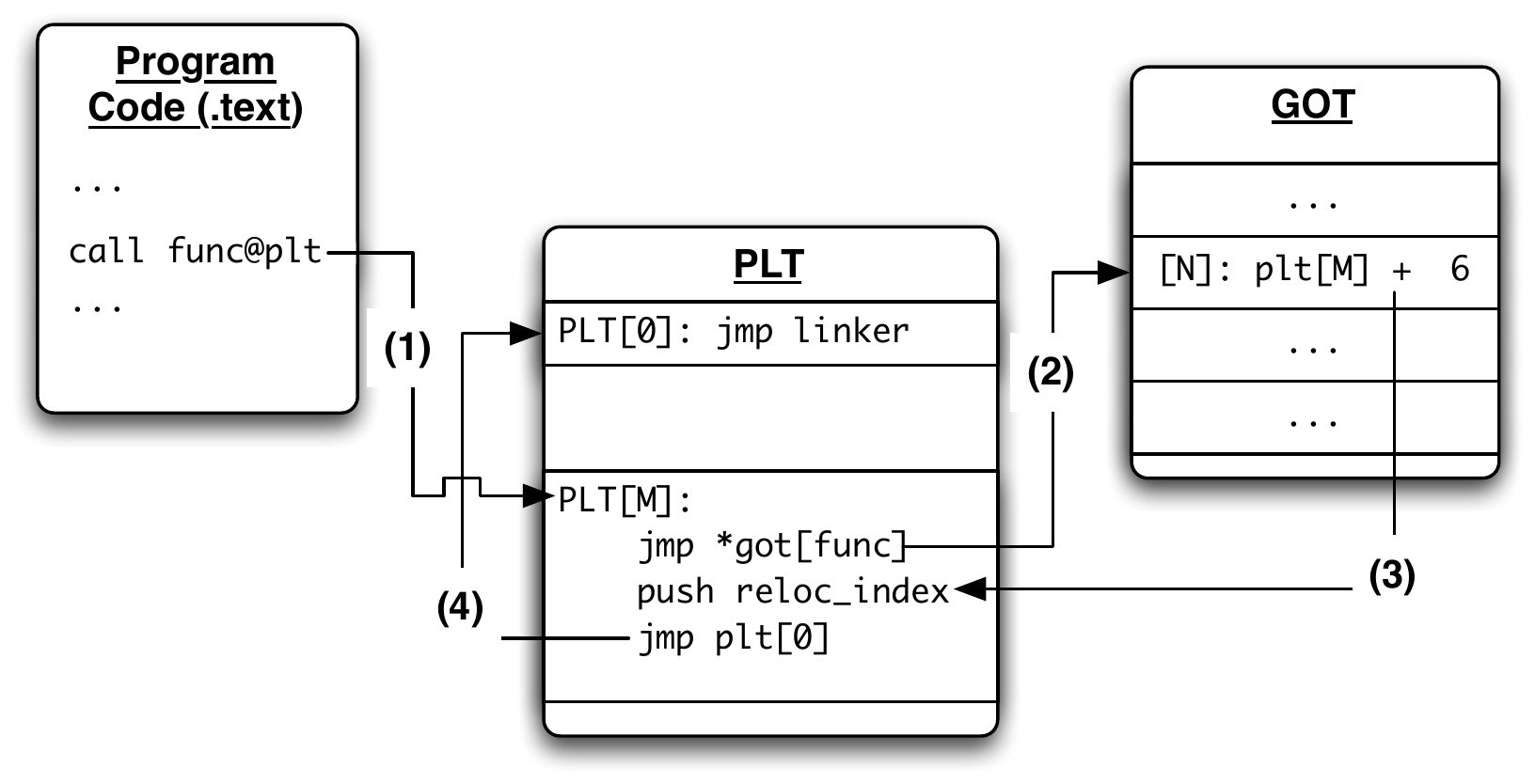}
          \caption{Function call symbol resolution using the Global
            Offset Table (GOT) and Procedure Linkage Table (PLT).}
          \label{fig:got-plt}
     \end{center}
\end{figure}

The \emph{Global Offset Table} (GOT) and \emph{Procedure Linkage
  Table} (PLT) are structures created by the runtime linker/loader
that can be manipulated by malware to provide execution in a victim
process without modifying the text segment. Both the PLT and the GOT
support dynamic library linking and position independent code
(PIC)\cite{elfspec}. The GOT holds the runtime addresses of global
data and functions that may not be known at compile time. The PLT
holds executable code used to make external function calls via the
GOT. Depending on how the program is linked, the GOT will be populated
with the correct function addresses either at program load time or on
the first invocation of each function.  Figure
\ref{fig:got-plt} shows how the PLT and GOT are used for late
binding of calls. (1) The program code calls into the PLT, (2) the PLT
jumps to the address stored in the GOT. If the function's address
hasn't been resolved yet, then (3) the address in the GOT is the next
instruction in the PLT (after the instruction that just jumped into
the GOT). (4) The PLT entry pushes information about the call target
onto the stack, then (5) jumps to the zeroth entry in the PLT to
invoke the dynamic linker. The dynamic linker resolves the address of
the desired function, stores it in the GOT, and executes the function.

Implants with the ability to modify a process's GOT or PLT can easily
redirect all invocations of a shared library function to trigger a
function in an executable region controlled by the attacker.

These techniques are slightly harder to detect than direct
modifications to a process's text section because the PLT and GOT
values may be unique per execution of a program. On the Intel
architecture PLTs reside in a shared text segment\cite{elfspec}, so
detecting a PLT hook is equivalent to detecting a text segment
modification. The GOT, however, is populated at load or run time, but
its contents are predictable given the base load addresses of the
process's segments. A procedure for detecting some GOT overwrites has
already been established for memory
forensics\cite[Chapter~25]{artofmemoryforensics}.

\subsection{Shared Object Injection}
Programs that support a plugin API often include the ability to load
arbitrary shared object files to extend their built-in
functionality. This can be directly exploited by implant authors to
introduce malicious functionality into a process. Shared object
injection is not strictly memory-only; it generally requires creation
of a file in the filesystem. Thus, implants based on shared object
injections can be detected by hashing the images of files as they are
mapped (as Linux's Integrity Measurement Architecture (IMA) does). In the unlikely event that shared objection
injection attacks are possible without modifying the filesystem in a
detectable way, these attacks are also detectable by validating the
set of files mapped by each process.

\emph{Pre-loading implants} are a common special case of shared object
injection. The runtime linker/loader supports specification of extra
shared libraries that should be mapped into every process based on the
\texttt{LD\_PRELOAD} environment variable or the contents of
\texttt{/etc/ld.so.preload}. By setting this variable, an attacker can
cause their implant to be mapped into arbitrary processes.

\subsection{Thread Injection}
Given the ability to run arbitrary code in a process, it's trivial
for an attacker to spawn a new thread via the \texttt{clone} system
call. This provides stealthy execution and gives the implant ongoing
access to the victim process's resources.

A maliciously injected thread of this nature is difficult to detect
because multi-threaded programs are common and threads don't carry
state explaining their genesis. However the code used to spawn
the thread may reside in a suspicious memory mapping (e.g., one with
both writeable and executable permissions or one that is anonymous
and executable) or the thread may execute code in a suspicious memory
mapping.

\subsection{Namespace Manipulation}
Namespaces provide isolation between processes with respect to a
global system resource. If two processes are in different namespaces,
they are invisible to each other with respect to the associated
resource. This feature is commonly used as a form of lightweight
virtualization called containers\cite{containers}. An adversary can
achieve stealthy execution by running an implant in different
namespaces from the rest of the system. The Horse Pill
rootkit\cite{horsepillpoc} implements this namespace isolation by
installing a custom initial RAM disk (\texttt{initrd}) image that executes the
rootkit functionality in the default namespaces and creates separate
namespaces for the normal system boot process\cite{horsepillanalysis}.

A namespace is identified by its type and inode number. These
inode numbers begin at hard-coded default values that are distinct for
each namespace type. If the \texttt{init} or \texttt{systemd} process
has non-default namespace inode numbers, then there may be processes
operating in different namespaces, outside the purview of most of the
userspace processes.

\subsection{File Descriptor Passing}
File descriptors can be passed between processes using UNIX domain
sockets via ancillary data\cite[Chapter~17]{fdpassing}. As such, an
attacker could inject into a process and copy or send file descriptors
to other processes. This could be used to bypass a firewall or port
binding restrictions (e.g., requiring superuser privileges to bind
well-known ports) or more generally to bypass access controls on any
kernel abstraction with a file interface.

As with thread injection there is nothing inherent to file descriptors
that could indicate malicious origins, but peripheral artifacts may
enable detection. For example if a file has attributes indicating
only the root user can access it, we do not expect a non-root process
to possess a file descriptor for that file. Similarly we might not
expect a process without superuser privileges to have a TCP or UDP
socket bound to a well-known port. More generally a process should not
have file descriptors for objects unless the process meets the access
requirements for those objects. Any deviation from this expectation is
suspect. Additionally the code that passed the file descriptor may
reside in a suspicious memory mapping in either the sending or
receiving process or in both.


\section{Implementation} 
\label{sec:implementation}

The USIM Toolkit consists of two components that work together to 
evaluate the integrity of userspace.  The first component is a collection
agent which gathers point-in-time information on both global and
per-process system state. Specifically it collects

\begin{itemize}
     \item System information: operating system, architecture, network
     name, and software inventory as reported by the native package
     manager
     \item Hashes of various important files on the system
     \item Meta-data for each process on the system
     \item Memory mappings, including permissions, addresses, and
     backing files and offsets for root-owned processes
     \item Namespaces in use on the system, and a map of which
     processes belong to which namespace
     \item The number, type, and owning process of each open file
     descriptor in all root-owned processes
     \item Per-process relocations (i.e. GOT/PLT entries) for each
     root-owned process
     \item Hashes of each executable memory segment currently mapped in
     a root-owned process
\end{itemize}

The choice to focus on root-owned processes is intended to limit the
performance impact of the USIM Toolkit while providing adequate
detection capabilities. This is a configuration option that can be
trivially changed to include measurement of all, or an expanded subset
of, processes. The USIM Toolkit is designed to be extensible, we
expect to incorporate additional tools for measuring new aspects of
system state.

\begin{figure}
     \begin{center}
          \includegraphics[width=5.5cm]{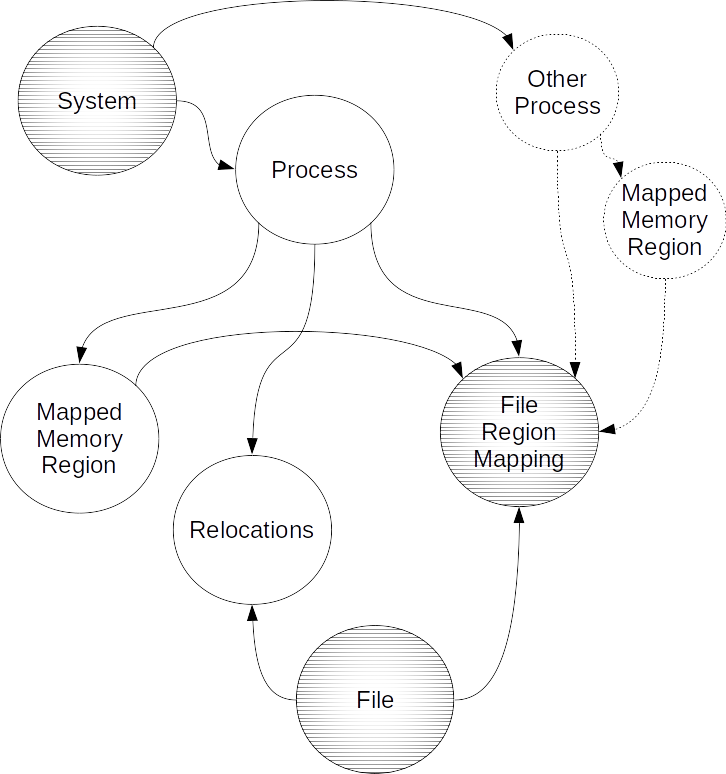}
          \caption{Example measurement graph subgraph showing the relationship between
            processes, process memory mappings, file memory mappings, and files.
            Shaded nodes can be used by multiple processes, while unshaded
            ones are unique to each process.}
          \label{fig:mgraph}
     \end{center}
\end{figure}

The collected information is gathered into a single graph-based data
structure which captures the complex relationship between the
individual data.  A graph also allows for multiple collection agents
to collect different sets of data in parallel, optionally with
different permissions.  The collection agent then bundles this
``measurement graph'' into a portable format for evaluation.  An
example subgraph showing memory-mapped regions of a process is shown
in Figure~\ref{fig:mgraph}.

The second USIM Toolkit component, called the appraiser, evaluates the
measurement graph to appraise the integrity of userspace.  The
appraiser extracts data and relations and evaluates the measurement
based on a set of rules in a policy defined by an administrator.  The
number and complexity of these rules is limited only by the resources
and data available to appraise. For this evaluation, our
implementation defines the following set of rules

\begin{enumerate}
     \item Only a defined subset of programs can have memory mappings
         that are both writable and executable (a whitelist)
         \item Only a defined subset of programs can arbitrarily 
         have file-backed executable memory mappings that do not
         correspond to a direct or transitive dependency of the
         binary (a whitelist)
     \item Only a defined subset of programs can have anonymous
         executable memory mappings (a whitelist)
     \item The original executable for a process must be in the
         file-backed mappings for the process
     \item Read-only executable memory mappings for processes should
         hash to the same values as the respective sections of the
         associated on-disk files
     \item The \texttt{init} or \texttt{systemd} process that is the
         ultimate parent of all other processes should be in the default
         PID namespace
     \item Resolved GOT entries should not change (e.g., the resolved
         address of \texttt{printf} should not change across
         snapshots once it has been resolved)
     \item A socket in use by one processes should not later be used
         in a different, non-child process
\end{enumerate}

The USIM appraiser determines whether the collected data conforms to
the defined policy and alerts the administrator with the
result. Separating the collection component from the appraisal piece
achieves two goals: 1) adherence to the principle of least privilege:
while the collection agent(s) may require elevated privileges to
collect data, the evaluation component often does not; and 2)
flexibility: the two components can be executed on separate machines,
and additional appraisal constraints can easily be added to more
tightly confine the allowable states of the measured platform.
Indeed, it is envisioned that in most scenarios the USIM evaluator
component will be run on a remote appraisal server, with only the
collection agent running on the client.

Although it should be possible to examine all of a process's GOT
entries and ascertain whether they point to reasonable locations, that
was beyond the scope of this work. For our purposes it was sufficient
to record GOT entries and manually detect changes across snapshots.
This does limit the utility of our approach to catching a GOT
infection as it happens rather than detecting any GOT infection, but
this was reasonable for our needs. A more thorough dynamic linking
integrity verification could simulate the dynamic linker's load
procedure, using the on-disk binary files, to verify that all observed
entries match the expected value.

Several of our appraisal rules use whitelists to broaden applicability
to a wider set of scenarios. For example interpreters, such as Python,
Java, and web browsers, often have benign anonymous executable memory
mappings or memory mappings that are both writable and executable. To
accommodate these cases, we relax those rules and adopt a whitelist to
provide flexibility. Additionally, some programs dynamically load
plugins which are not explicitly listed as dependencies
(e.g. apache2), and we address this by using another whitelist. For
this work we na\"{\i}vely whitelisted by binary executable name, and
the whitelists were not implemented in the USIM Toolkit.  Rather the
USIM Toolkit reports the suspicious observations and the onus is on
the administrator to render the final appraisal decision.

We implemented both components of the USIM Toolkit using plain C99
language constructs with minimal external library dependencies. The
collection agent is implemented as a collection of separate programs
that are executed by a central control process. This allows each
program to be run with a minimal set of privileges (e.g., the file
hashing program does not require the ability to read arbitrary
process's memory). The USIM Toolkit targets the GNU/Linux operating system,
but could likely be retargeted to other operating systems by porting
the relevant collection subprograms.

Rules 7 and 8 also require access to previous measurements to identify
changes over time. As of this writing, our implementation is not yet able
to automatically compare current results to previous results, but the
feature is in progress. Therefore, in this paper, we manually verify 
successive reports to verify those parts of policy. While
not ideal, the fact that such verification can be done with the
collected information is sufficient to evaluate the utility of the
USIM Toolkit.


\section{Evaluation}
\label{sec:evaluation}
Security solutions must be capable of detecting malicious activities
without imposing unacceptable performance overhead. This section
presents a functional evaluation of the USIM Toolkit based on its
ability to detect proof-of-concept implementations of the implant
techniques described in Section \ref{sec:implant-techniques}, and
initial benchmarks of the performance overhead of the USIM Toolkit.

\subsection{Functional Evaluation}
To evaluate the effectiveness of the USIM Toolkit in detecting common
implant techniques, we implemented representative samples of the
techniques described in Section \ref{sec:implant-techniques} and
observed the detection capability of the USIM Toolkit. Our original
goal was to evaluate the USIM Toolkit against implant samples
collected from the Internet. Unfortunately, running such an evaluation
against real implants is extremely challenging because (a) implants
may be highly tuned to a specific runtime environment and command and
control system that is difficult to replicate in a lab environment,
(b) running real implants even in a controlled environment risks an
outbreak that could damage operational networks, and (c) individual
implants may combine multiple known and unknown techniques which makes
it difficult to test specific detection hypotheses. Based on these
limitations, we opted to develop clean-room implementations of
specific implant techniques and test detection of these samples
instead. We used open-source implementations for any techniques where
straightforward implementations with no additional functionality were
readily available. Notably for process text segment modification and
name\-space manipulation we used Modern Userland
Exec\cite{userlandexec} and Horse Pill\cite{horsepillpoc},
respectively.

The USIM Toolkit is not intended to detect initial process
exploitation. Accordingly, our implant samples do not include first
stage exploit capabilities such as buffer overflows or ROP chains. The
samples are simple C programs that each demonstrate a particular
technique, excepting the two open-source samples. The narrow scope of
each sample made it simple to isolate and test detection of the
implant behavior under consideration. For similar reasons, we also
disabled the built-in exploit mitigation functions such as address
space layout randomization, stack canaries, $W \oplus X$ memory
protections, and the SELinux mandatory access control system. These
mechanisms are an important part of host defenses, but measurement
systems like the USIM Toolkit provide an important fallback to detect
when protections fail.

\begin{table}
        \begin{tabular}{l c } 
          \toprule
          Implant Class & Rules Triggered \\
          \midrule
          Text Segment Mod & 1, 4, 5 \\
          GOT/PLT Hooking & 7 \\
          Shared Object Injection & 2 \\
          Thread Injection & 1, 3, 4 \\
          Namespace Manipulation & 6 \\
          FD Passing & 8 \\
          \bottomrule
        \end{tabular}
        \caption{Summary of effectiveness against each class of implants,
                as well as whether other known techniques would catch
                the implant.}
        \label{tab:effective}
\end{table}

\subsubsection{Process Text Segment Modification}
We used the open-source tool Modern Userland Exec\cite{userlandexec} to
evaluate text segment modification. This tool is especially interesting
because instead of merely modifying part of the text segment, it mimics
the \texttt{execve} system call and completely replaces the original
process's text segment with that of another binary. However, unlike
\texttt{execve}, this is done with anonymous memory mappings and without
causing changes to most of the kernel's information about the process
(e.g., \texttt{/proc/[pid]/exe}). This allows one program to masquerade
as another.

The USIM Toolkit appraiser alerted on this process for two reasons: 1)
the process had anonymous executable memory mappings and the program
is not on the whitelist of interpreters, and 2) the initial executable
was missing from the file-backed mappings.

\subsubsection{GOT/PLT Hooking}
We evaluated GOT and PLT hooking by authoring a simple C program that
prompts the user for a command. The user supplies either ``got'' or
``plt'' and then an address. For the ``got'' command, the address is the
address of the GOT entry for \texttt{\_\_xstat}, and the program
overwrites that entry with the address of a malicious function. Then
the program calls \texttt{stat} to demonstrate the hook.

For the ``plt'' command, the user-supplied address is the address of the
PLT entry for \texttt{\_\_lxstat}, and the program uses
\texttt{mprotect} to make that page of memory writable, overwrites the
first instruction of that PLT entry with a call to a malicious
function, and then uses \texttt{mprotect} to make the page read-only
again. This solution is less portable than the GOT solution because it
relies on the first instruction at that PLT address being a 6-byte
jump and overwrites it with a 5-byte call to a malicious function and
a 1-byte no-op. This also has the side effect of forcing the dynamic
linker to resolve the address of \texttt{\_\_lxstat} every time
\texttt{lstat} is called. The program then calls \texttt{lstat} to
demonstrate the hook.

This program requires disabling RELocation Read-Only
(RELRO)\cite{relro}. Otherwise the GOT would be unwritable, preventing
our GOT hook, and the dynamic linker's reserved GOT entries would be
zeroed out, causing our PLT hook to crash the program. Our PLT hooking
technique was to overwrite the jump instruction with a function call
and a no-op in order to call our malicious function and then rely on
the PLT's symbol resolution to execute the intended function.
This symbol resolution involves jumping to the address stored in the
third entry of the GOT, which would normally be the address of the
dynamic linker's symbol resolution function. However on our Ubuntu
17.10 system, RELRO binaries are missing the \texttt{.got.plt} section
and the second and third entries of the GOT are zeroed out. Thus,
jumping to the address in the third GOT entry causes a segmentation
fault. More sophisticated GOT and PLT hooks could easily bypass this
defense mechanism by making the GOT writeable before writing to it
(and then making it read-only again for stealth) and by using a
PLT hook that preserves the PLT's behavior (i.e. jumping to the address
in the GOT and triggering symbol resolution if the symbol hasn't been
resolved yet).

The USIM Toolkit appraiser detected the GOT modification via 
resolving each relocation both before and after the infection, and
noting the change.  In our case, the GOT entry for
\texttt{\_\_lxstat} pointed to the binary's text segment mapping
rather than a mapping to libc. Similarly, the appraiser
detected the PLT modification because the hash of that memory mapping
no longer matched its on-disk representation in the binary, as well as
detecting that change in relocation value from before the infection.

\subsubsection{Shared Object Injection}
For shared object injection, we composed a simple C program that takes
the path to a shared object and the name of its entry point symbol,
calls \texttt{dlopen} to load the shared object, uses \texttt{dlsym} to
find the entry point, and then calls the entry point.

The loaded shared object was not one of the binary's dependencies, and was
located outside of the normal library paths. The USIM Toolkit appraiser 
detected this, as this binary was not whitelisted to allow arbitrary loading
of shared objects. 

\subsubsection{Thread Injection}
We achieved thread injection via shellcode that uses the
\texttt{clone} system call to create a new thread whose entry point is
elsewhere in the shellcode. We fed the shellcode to a simple program
that reads network data into a buffer, marks the buffer as executable,
and calls into it. We wrote the shellcode in x86\_64 assembly that we
adapted from an open-source example\cite{purelinuxthreadsdemo}.

While the USIM Toolkit appraiser had no way of knowing how many threads
each process should have, it was able to notice that the
parasite thread was executing code in an anonymous executable mapping,
which our restrictive appraisal policy forbids for processes not in the 
whitelist, so the system failed appraisal.

\subsubsection{Namespace Manipulation}
We used Horse Pill\cite{horsepillpoc}, an open-source implementation of
a custom initrd, to evaluate namespace manipulation. The custom initrd
has a malicious \texttt{run-init} that compromises a system on boot.
The modified \texttt{run-init} migrates most of the system to a new PID
namespace, makes fake kernel thread processes in that namespace, and
runs a backdoor outside the namespace.

Even running within Horse Pill's infection namespace, the USIM Toolkit
was able to gather sufficient evidence to cause the appraisal to fail
because the init process's namespace is not the expected default.  Also, the 
total number of namespaces on the system differed from the expected value.

\subsubsection{File Descriptor Passing}
To evaluate file descriptor passing, we authored a simple C program that
accepts a TCP connection and sends it to another process via UNIX domain
socket.

Two subsequent USIM Toolkit measurements showed the same socket file
descriptor was open by one process in the first measurement, and another
process in the second measurement.  Manual inspection showed that
the information was present to accurately identify and fail the second
appraisal.

\subsection{Performance Evaluation}
\label{sec:performance-evaluation}

\begin{table*}[ht]
  \scriptsize
  \begin{tabularx}{\textwidth}{X X  X  X }
       \toprule
       & CentOS 7\newline (VM, Idle) & CentOS 7\newline (VM, Load) & Ubuntu 17.10\newline (HW, Desktop) \\
       \midrule
    Graph Nodes     & 6275  & 6750 & 14804 \\ 
    Graph Edges     & 27837 & 30275 & 14804 \\
    Total Size (MB) & 12.5  & 13.3 & 30.3 \\
    Processes       & 123   & 151 & 341 \\
    Root-Owned Processes &   27 & 47 & 54 \\
    Collection Time (s) & 199 & 606 & 255 \\
    Per-Process Time (s) & 7.5 & 16 & 4.6 \\
    Peak CPU Usage (\%) & 100 & 100 & 100 \\
    Peak Mem Usage (MB) & 67 & 67 & 80 \\ 
    Appraisal Time (s) & 649 & 1575 & 1594 \\
    \bottomrule
  \end{tabularx}
  \caption{USIM measurement metrics in three test scenarios. Average of 10 measurements.}
  \label{table:benchmarks}
\end{table*}

We measured the performance overhead of the USIM Toolkit by performing
complete measurements of (a) a freshly booted minimal CentOS 7 virtual
machine, (b) A CentOS 7 VM under a synthetic load of benchmarking 
redis server performance while simultaneously compiling a Linux kernel, 
and (c) a typical Ubuntu 17.10 user desktop system after
days of typical user activity (web browsing, email, etc). Systems (a)
and (b) are virtual machines using the KVM hypervisor on system (c),
with 4 virtual CPUs and 4 GB of memory assigned to them, while system 
(c) is a Dell Latitude E7450 laptop with 16GB of memory and an Intel i7-5600U 
CPU running at 2.6 Ghz.  All appraisals were processed on the Dell E7450
laptop. 
 
For each case we collected the total CPU time, peak CPU usage, peak 
memory usage, and network data transferred. Table \ref{table:benchmarks} lists
average values over 10 runs for these each of the three test scenarios.

We also measured the effect of measurement on the performance of a
system under heavy load.  We collected benchmark data from the 
\texttt{redis-benchmark} tool on an idle CentOS 7 VM with 4 vCPUs, the 
CentOS 7 VM while compiling the Linux kernel with \texttt{make -j4} 
with an \textit{allyesconfig} configuration, and when taking a USIM measurement 
with the above load. This load ensured that the system was handling heavy disk,
CPU, memory, and network load at the time of measurment.  We repeated the 
benchmark 12 times for each condition and computed the averages by throwing
away the max and minimum results and averaging the remaining 10 for each
redis benchmark subtest.  The results of this experiment are shown in 
Table~\ref{tab:redis}.  

\begin{table*}[h!]
\scriptsize
\begin{tabularx}{\textwidth}{l X X X X X X}
\toprule          
Redis benchmark	& Idle
                & USIM
                & Delta
                & Load
                & Load+USIM
                & Delta \\
                & (\% of Idle)
                & (\% of Idle)
                & (Idle-USIM)
                & (\% of Idle)
                & (\% of Idle)
                & (Load - \newline (Load+USIM)) \\
                
\midrule
GET	       & 100.00	& 74.42	& 25.58	& 30.97	& 23.66	& 7.30 \\
HSET           & 100.00	& 72.89	& 27.11	& 29.01	& 20.40	& 8.61 \\
INCR           & 100.00	& 73.12	& 26.88	& 31.09	& 21.95	& 9.14 \\
LPOP           & 100.00	& 71.37	& 28.63	& 29.19	& 22.44	& 6.75 \\
LPUSH          & 100.00	& 71.87	& 28.13	& 29.53	& 23.42	& 6.12 \\
LRANGE\_100    & 100.00	& 72.27	& 27.73	& 28.88	& 22.56	& 6.32 \\
LRANGE\_300    & 100.00	& 68.45	& 31.55	& 26.38	& 20.25	& 6.13 \\
LRANGE\_500    & 100.00	& 75.82	& 24.18	& 30.99	& 23.70	& 7.28 \\
LRANGE\_600    & 100.00	& 70.72	& 29.28	& 30.93	& 23.90	& 7.03 \\
MSET (10 keys) & 100.00	& 69.41	& 30.59	& 31.70	& 24.47	& 7.23 \\
PING\_BULK     & 100.00	& 64.88	& 35.12	& 26.52	& 19.50	& 7.02 \\
PING\_INLINE   & 100.00	& 75.69	& 24.31	& 32.33	& 24.20	& 8.13 \\
RPOP           & 100.00	& 74.72	& 25.28	& 33.36	& 24.87	& 8.48 \\
RPUSH          & 100.00	& 73.46	& 26.54	& 30.17	& 23.32	& 6.86 \\
SADD           & 100.00	& 72.85	& 27.15	& 29.38	& 21.98	& 7.40 \\
SET            & 100.00	& 74.69	& 25.31	& 28.84	& 19.86	& 8.98 \\
SPOP           & 100.00	& 74.11	& 25.89	& 30.50	& 22.26	& 8.24 \\
\midrule                                                 
Avg            & 100.00	& 72.40	& 27.60	& 29.99	& 22.51	& 7.47 \\
\bottomrule
\end{tabularx}
\caption{Redis benchmark performance degradataion attributable to the
  USIM Toolkit relative to an idle system (larger is better)}
\label{tab:redis}
\end{table*}

These results show a significant impact of our un-optimized protoype on the 
Redis benchmark in isolation.  Initial tests on an otherwise idle system show
on average 28\% less performance when taking a measurement. However, when
the system is under stress from another workload, the additional impact of
USIM on the original benchmark is only 8\%. The significantly lower impact on 
the loaded score suggests that much of the performance impact of measurement is 
related to running any task in addition to the benchmark suite.

While some impact on the system is unavoidable, these benchmarks suggest that 
significant work is needed to reduce the performance impact of the USIM Toolkit.
Implementing local
evaluation checks to limit the amount of data that must be cached and
sent to the appraiser will likely greatly reduce the peak memory usage
and network IO of the USIM Toolkit. When implementing the USIM
Toolkit, we paid little attention to performance optimization, so many
implementation optimizations are likely available that would reduce
the CPU time and memory required.


\section{Related Work}
\label{sec:related-work}
Computer defense has long been a ``cat and mouse game'' in which
attackers continually discover and deploy new mechanisms for
exploiting and controlling their victims, and defenders respond by
developing counter measures to prevent exploitation or detect the
control mechanism.  Defensive strategies include active mitigations,
runtime integrity measurement, and hardware-based trust mechanisms, as
well as traditional antivirus and enterprise client management
tools. Defenders use these tools to attempt to prevent 
adversaries from gaining access to systems, to detect when a system
has been compromised, and to limit the effects a compromise might have
on a single host or across a network.

\subsection{Active Mitigations}
Many defensive measures, such as $W \oplus X$\cite{wxorx} and Address
Space Layout Randomization (ASLR)\cite{ASLR}, introduce challenges to
initial exploitation, but experience has shown that adversaries are
able to work around these mitigations, using techniques such as memory
disclosures, information side-channels, and return-oriented
programming\cite{memory-disclosures, information-sidechannels,
  rop,hacking-blind, rop-still-dangerous},
to install implants for long-term control. Operating system controls
such as access-control based sandboxing\cite{sandbox, selinux} and code
signing attempt to mitigate implants by limiting what executables can
be run from different security contexts, but various attacks have
shown that these too are circumvented by
adversaries\cite{chrootjail-breakout, sysjail-breakout, signed-malware}.

\subsection{Measurement Agents}
Integrity measurement tools attempt to detect implants by attesting to
the integrity of a system.  A key advantage for implant detection
tools is that implants intentionally have lasting effects on a victim
platform such as providing a command and control channel for the
adversary. Measurement tools, such as Linux's Integrity Measurement
Architecture (IMA)\cite{ima}, the Linux Kernel Integrity Measurer
(LKIM)\cite{lkim}, and Semantic Integrity\cite{semantic-integrity},
attempt to identify these effects on either the filesystem or the
runtime behavior of processes. These existing integrity measurement
solutions have focused on pre-boot environments, the operating system
kernel, or file images. Userspace integrity measurement extends the
concepts introduced in these works to support verification of the next
abstraction level in a modern system, the userspace operating
environment. Comprehensive system integrity measurement should
include application-level measurements, and runtime measurements of
lower platform levels such as the hypervisor or system management
mode.

\paragraph{Dynamic Measurements}
LKIM\cite{lkim} and Semantic Integrity\cite{semantic-integrity} are
dynamic measurement techniques that can measure kernel data structures
at any time during a platform's execution. Like the USIM Toolkit, these
tools work by inspecting their target's runtime state to identify
violations of key invariants that may indicate compromise.  Unlike the
USIM Toolkit, these tools focus on implants that operate by modifying
data structures in the kernel's memory space. Combining these kernel
integrity measurement solutions with userspace integrity measurement
can significantly reduce the opportunities for an implant to hide in
a modern system. Filling in the gaps, for example by adding
measurement of the USIM Toolkit to a kernel-level solution, is an
important area of future research.

\paragraph{Static Measurements}
Static measurement tools, such as IMA and Cb Protection, attempt to
guarantee integrity by taking a cryptographic hash of files at
loadtime. IMA is functionality built into the Linux kernel that
performs cryptographic hashes of all, or a configured subset of, files
accessed by a system. The log of these hashes can be reported to
userspace along with a certification of the current value using a
Trusted Platform Module (TPM). IMA builds on a long history of
file-hashing based approaches to system integrity validation;
TripWire\cite{tripwire} is the earliest notable ancestor of IMA. IMA
improves on historic approaches primarily by (a) hashing files as they
are accessed, (b) performing the hash from within the operating system
kernel context, and (c) using a TPM to endorse the hash log. To detect
implants using a system like IMA, a trusted system can compare the
reported log with either a blacklist of known implants or a whitelist
of approved files. Cb Protection from Carbon Black
\cite{carbon-black-protection} also provides load-time checks on
programs as they are launched in order to verify integrity at program
start.

In environments, such as fixed-purpose embedded systems, where a
complete whitelist of valid files is tractable, static measurements
can provide strong guarantees that no implants are loaded from a
system's filesystem. Unfortunately, most systems are not amenable to
complete whitelists, and creation of a comprehensive blacklist of
malware hashes is infeasible because new variants with new hashes are
trivially created by attackers. Further, memory-only implants may not
require loading any attacker-modified files from the filesystem. This
paper complements these approaches by extending integrity from the
filesystem to other key aspects of the userspace runtime environment
such as environment variables, interprocess communication channels,
and runtime linker-loader behavior.

\paragraph{Root of Trust} 
Confidence in measurement tools requires a root of trust to guarantee
that the measurements are correctly collected and reported. In their
research, England et al.\cite{trusted-open-platform} describe an
approach to integrity measurement in which measurements are
cryptographic hashes of boot-time software. This approach is also used
by the Trusted Computing Group (TCG) as part of the ``Trusted Boot''
technology\cite{tboot}. SecureBoot \cite{secure-boot} also aims to
provide a chain of integrity checks beginning at system power-on. These 
checks are optimal for identifying persistant threats on the platform, but 
are largely ineffective against attacks that take place after startup. 
The work described in this paper aims to extend this trusted base to
userspace software at runtime in order to more effectively guard
against a wider range of attacks.

\subsection{Antivirus}
Integrity measurement attempts to enforce a set of invariants to which
any well-behaved system should conform. Antivirus tools take the
opposite approach, defining specific static or behavioral signatures
that only malicious software should exhibit. The specific techniques
used by antivirus products are generally proprietary, but most appear
to be based on fingerprinting files and monitoring runtime process
behavior such as system call tracing\cite{antivirus, behavioral-detection}. Fingerprinting
files may be more resilient to variations than the precise
cryptographic hashes used by IMA, but decades of experience has shown
that implant authors are able to quickly adapt and deploy variants
with previously unknown
signatures\cite{malware-obfuscation, hidden-malware}.

Behavioral profiling is another powerful tool that has proven
useful in recognizing many implants as they are executing. Behavioral
approaches are enhanced by the ability of antivirus vendors to collect
large scale data from across their customer base and correlate newly
observed behavior with historical indicators of infections. This is
most effective in detecting broadly distributed implants that may
trigger other alerting systems such as network monitoring solutions.

Some antivirus tools, such as Cb Defense \cite{carbon-black-defense}
have begun to use predictive modeling to identify when a process is
behaving suspiciously. However, as is often the case in the arms race 
for cyber security, researchers are already inventing ways to hide their
malicious behavior in the execution of benign processes \cite{malwash}.

\paragraph{Enterprise Defense}
Enterprise defense frameworks have similar goals to traditional
antivirus tools, but at a larger scale.  These include Trusted Network
Connect (TNC) and SAMSON. Similar to the limitations with antivirus
tools, these frameworks have a narrow focus and would generally not
detect the attacks evaluated in this paper. Specifically, TNC is
focused on attestation only within the scope of access
control\cite{TNC}, while SAMSON is designed to do remote attestation
of client systems using IMA logic to gauge the integrity of programs
running on the system\cite{samson}.  Neither of these frameworks
introduces novel measurements capable of detecting the userspace
attacks targeted by this paper.

More recent enterprise defense products are incorporating integrity
measurement-like functionality. Forcepoint Threat Protection for
Linux has claimed success in detecting Horse
Pill\cite{horsepillanalysis}. Now known as Forcepoint Linux Security
or Second Look, this tool uses memory forensics to identify malicious
software on the system. The backend of this tool is proprietary;
however, it seems to rely on comparing the captured kernel memory
image to a reference kernel, and taking pagewise hashes of executables
to compare to known-good versions. These abilities would make it
effective against some of the implant techniques listed in this
paper.


\section{Future Work}
\label{sec:future-work}
The USIM Toolkit includes measurements of process environments,
runtime structures, and access to OS resources. We evaluated these
measurements against proof-of-concept implementations of common
implant techniques to demonstrate that they are able to reliably
detect implants that may elude detection by traditional means. It is
unlikely that these measurements are comprehensive; continued study of
userspace invariants and how they are violated by advanced malware is
an important area of continued research.

Performance evaluation of the USIM Toolkit showed a significant
performance impact, a full measurement of an active system took
over 200 seconds, required 80 MB or more of memory, and transferred
20+ MB of data to a remote appraiser. We believe that USIM
Toolkit prototype implementation could be significantly optimized, the
data collected for the same measurements could be reduced, and some
local checks could be performed to minimize the data transmitted to
the remote appraiser. Implementing these improvements in order to
reduce the performance overhead is important to ensure these
approaches are usable in practical contexts.

Userspace measurement is only one part of producing comprehensive
integrity measurement of a modern platform. Prior work largely focuses
on kernel-level measurements. Future work may introduce strategies for
measurements of individual security-critical applications and measurement
of lower-level components such as hypervisors. To fully benefit from
these components, significant additional work must also be done to
understand how measurements can be combined to ensure that the correct
appraisal of a lower-level measurement justifies trust in higher-level
measurements.


\section{Conclusion}
\label{sec:conclusion}

This paper introduced the USIM Toolkit, an extensible set of GNU/Linux
measurement and appraisal tools for verifying the integrity of
userspace. We have shown that this prototype is capable of detecting a
variety of sophisticated implant techniques. Although work is needed to
improve the toolkit's completeness and performance, it is a general
mechanism to detect a broad class of integrity violations with myriad
security implications. Because the USIM Toolkit is based on invariants
of well-behaved systems, it is part of a workable integrity strategy
that requires no preknowledge of specific attacks. By combining the USIM
Toolkit with other integrity verification techniques, trust could be
extended from a root of trust to the application level to form a
comprehensive verification solution. Even on its own the USIM
Toolkit is a critical advancement in the detection of sophisticated
in-memory userspace implants.


\Urlmuskip=0mu plus 1mu\relax\raggedright
\bibliography{bibliography}{}

\begin{thebibliography}{10}

\bibitem{carbon-black-defense}
Cb defense.
\newblock \url{https://www.carbonblack.com/products/cb-defense/}, 2018.
\newblock Accessed: 2018-03-07.

\bibitem{carbon-black-protection}
Cb protection.
\newblock \url{https://www.carbonblack.com/products/cb-protection/}, 2018.
\newblock Accessed: 2018-03-07.

\bibitem{secure-boot}
W.~A. Arbaugh, D.~J. Farber, and J.~M. Smith.
\newblock A secure and reliable bootstrap architecture.
\newblock In {\em Security and Privacy, 1997. Proceedings., 1997 IEEE Symposium
  on}, pages 65--71. IEEE, 1997.

\bibitem{hacking-blind}
A.~Bittau, A.~Belay, A.~Mashtizadeh, D.~Mazières, and D.~Boneh.
\newblock Hacking blind.
\newblock In {\em 2014 IEEE Symposium on Security and Privacy}, pages 227--242,
  May 2014.

\bibitem{unix-rootkits}
A.~Bunten.
\newblock Unix and linux based rootkits techniques and countermeasures.
\newblock In {\em 16th Annual First Conference on Computer Security Incident
  Handling, Budapest}, 2004.

\bibitem{rop-still-dangerous}
N.~Carlini and D.~Wagner.
\newblock {ROP} is still dangerous: Breaking modern defenses.
\newblock In {\em 23rd {USENIX} Security Symposium ({USENIX} Security 14)},
  pages 385--399, San Diego, CA, 2014. {USENIX} Association.

\bibitem{chrootjail-breakout}
A.~Chuvakin.
\newblock Using chroot securely.
\newblock \url{http://www.linuxsecurity.com/content/view/117632/49/}, November
  2007.
\newblock [Online; accessed: 2018-02-28].

\bibitem{crowdstrikecasebook}
CrowdStrike.
\newblock Crowdstrike releases annual cyber intrusion services casebook.
\newblock
  \url{https://www.crowdstrike.com/resources/news/crowdstrike-releases-annual-cyber-intrusion-services-casebook/},
  December 2017.
\newblock [Online; accessed: 2018-03-12].

\bibitem{wxorx}
T.~de~Raadt.
\newblock W\^{}{X} - the mechanism.
\newblock \url{http://www.openbsd.org/papers/ven05-deraadt/mgp00009.html}, May
  2006.
\newblock [Online; accessed: 2018-02-26].

\bibitem{tboot}
Dice.
\newblock Trusted boot (tboot).
\newblock \url{http://tboot.sourceforge.net}, 2015.
\newblock Accessed: 2015-03-20.

\bibitem{vmsvscontainers}
R.~Dua, A.~R. Raja, and D.~Kakadia.
\newblock Virtualization vs containerization to support paas.
\newblock In {\em 2014 IEEE International Conference on Cloud Engineering},
  pages 610--614, March 2014.

\bibitem{userlandexec}
B.~Edinger.
\newblock Modern userland exec.
\newblock \url{http://stratigery.com/userlandexec.html}, 2014.
\newblock [Online; accessed: 2018-01-29].

\bibitem{trusted-open-platform}
P.~England, B.~Lampson, J.~Manferdelli, M.~Peinado, and B.~Willman.
\newblock A trusted open platform.
\newblock {\em Computer}, 36(7):55--62, July 2003.

\bibitem{samson}
C.~Fisher, D.~Bukovick, R.~Bourquin, and R.~Dobry.
\newblock Samson - secure authentication modules.
\newblock \url{http://sourceforge.net/p/secureauthentic/wiki/Home/}, 2015.
\newblock Accessed: 2015-04-02.

\bibitem{malwash}
K.~K. Ispoglou and M.~Payer.
\newblock malwash: Washing malware to evade dynamic analysis.
\newblock In {\em 10th {USENIX} Workshop on Offensive Technologies ({WOOT}
  16)}, Austin, TX, 2016. {USENIX} Association.

\bibitem{behavioral-detection}
G.~Jacob, H.~Debar, and E.~Filiol.
\newblock Behavioral detection of malware: From a survey towards an established
  taxonomy.
\newblock {\em Journal in Computer Virology}, 4(3):251--266, Aug 2008.

\bibitem{relro}
J.~Jelinek.
\newblock [rfc patch] little hardening dsos/executables against exploits.
\newblock \url{https://www.sourceware.org/ml/binutils/2004-01/msg00070.html},
  January 2004.
\newblock [Online; accessed: 2018-01-11].

\bibitem{signed-malware}
D.~Kim, B.~J. Kwon, and T.~Dumitra\c{s}.
\newblock Certified malware: Measuring breaches of trust in the windows
  code-signing pki.
\newblock In {\em Proceedings of the 2017 ACM SIGSAC Conference on Computer and
  Communications Security}, CCS '17, pages 1435--1448, New York, NY, USA, 2017.
  ACM.

\bibitem{tripwire}
G.~H. Kim and E.~H. Spafford.
\newblock The design and implementation of tripwire: A file system integrity
  checker.
\newblock In {\em Proceedings of the 2Nd ACM Conference on Computer and
  Communications Security}, CCS '94, pages 18--29, New York, NY, USA, 1994.
  ACM.

\bibitem{horsepillpoc}
M.~Leibowitz.
\newblock Horse pill.
\newblock \url{https://github.com/r00tkillah/HORSEPILL}, 2016.
\newblock [Online; accessed: 2018-01-29].

\bibitem{lkim}
P.~A. Loscocco, P.~W. Wilson, J.~A. Pendergrass, and C.~D. McDonell.
\newblock Linux kernel integrity measurement using contextual inspection.
\newblock In {\em Proceedings of the 2007 ACM workshop on Scalable trusted
  computing}, pages 21--29. ACM, 2007.

\bibitem{tpm}
T.~Morris.
\newblock Trusted platform module.
\newblock In {\em Encyclopedia of Cryptography and Security}, pages 1332--1335.
  Springer, 2011.

\bibitem{hidden-malware}
P.~OKane, S.~Sezer, and K.~McLaughlin.
\newblock Obfuscation: The hidden malware.
\newblock {\em IEEE Security Privacy}, 9(5):41--47, Sept 2011.

\bibitem{ASLR}
{PaX Team}.
\newblock Address space layout randomization.
\newblock \url{https://pax.grsecurity.net/docs/aslr.txt}, March 2003.
\newblock [Online; accessed: 2018-02-26].

\bibitem{maat}
J.~A. Pendergrass, S.~Helble, J.~Clemens, and P.~Loscocco.
\newblock Maat: A platform service for measurement and attestation.
\newblock {\em arXiv preprint arXiv:1709.10147}, 2017.

\bibitem{semantic-integrity}
N.~L. Petroni~Jr, T.~Fraser, A.~Walters, and W.~A. Arbaugh.
\newblock An architecture for specification-based detection of semantic
  integrity violations in kernel dynamic data.
\newblock In {\em Usenix Security}, 2006.

\bibitem{bundling}
P.~D. Rowe.
\newblock Bundling evidence for layered attestation.
\newblock In M.~Franz and P.~Papadimitratos, editors, {\em Trust and
  Trustworthy Computing}, pages 119--139, Cham, 2016. Springer International
  Publishing.

\bibitem{rutkowska}
J.~Rutkowska.
\newblock Introducing stealth malware taxonomy.
\newblock
  \url{https://blog.invisiblethings.org/papers/2006/rutkowska_malware_taxonomy.pdf},
  November 2006.
\newblock [Online; accessed: 2017-11-20].

\bibitem{ima}
R.~Sailer, X.~Zhang, T.~Jaeger, and L.~van Doorn.
\newblock Design and implementation of a tcg-based integrity measurement
  architecture.
\newblock In {\em Proceedings of the 13th Conference on USENIX Security
  Symposium - Volume 13}, SSYM'04, pages 16--16, Berkeley, CA, USA, 2004.
  USENIX Association.

\bibitem{information-sidechannels}
J.~Seibert, H.~Okhravi, and E.~S\"{o}derstr\"{o}m.
\newblock Information leaks without memory disclosures: Remote side channel
  attacks on diversified code.
\newblock In {\em Proceedings of the 2014 ACM SIGSAC Conference on Computer and
  Communications Security}, CCS '14, pages 54--65, New York, NY, USA, 2014.
  ACM.

\bibitem{rop}
H.~Shacham.
\newblock The geometry of innocent flesh on the bone: Return-into-libc without
  function calls (on the x86).
\newblock In {\em Proceedings of the 14th ACM Conference on Computer and
  Communications Security}, CCS '07, pages 552--561, New York, NY, USA, 2007.
  ACM.

\bibitem{containers}
S.~Soltesz, H.~P\"{o}tzl, M.~E. Fiuczynski, A.~Bavier, and L.~Peterson.
\newblock Container-based operating system virtualization: A scalable,
  high-performance alternative to hypervisors.
\newblock In {\em Proceedings of the 2Nd ACM SIGOPS/EuroSys European Conference
  on Computer Systems 2007}, EuroSys '07, pages 275--287, New York, NY, USA,
  2007. ACM.

\bibitem{selinux}
R.~Spencer, S.~Smalley, P.~Loscocco, M.~Hibler, D.~Andersen, and J.~Lepreau.
\newblock The flask security architecture: System support for diverse security
  policies.
\newblock In {\em Proceedings of the 8th Conference on USENIX Security
  Symposium - Volume 8}, SSYM'99, pages 11--11, Berkeley, CA, USA, 1999. USENIX
  Association.

\bibitem{fdpassing}
W.~R. Stevens and S.~A. Rago.
\newblock {\em Advanced Programming in the UNIX Environment, Third Edition}.
\newblock Addison-Wesley Professional, 2013.
\newblock [Online] Available: Safari e-book.

\bibitem{memory-disclosures}
R.~Strackx, Y.~Younan, P.~Philippaerts, F.~Piessens, S.~Lachmund, and
  T.~Walter.
\newblock Breaking the memory secrecy assumption.
\newblock In {\em Proceedings of the Second European Workshop on System
  Security}, EUROSEC '09, pages 1--8, New York, NY, USA, 2009. ACM.

\bibitem{antivirus}
O.~Sukwong, H.~Kim, and J.~Hoe.
\newblock Commercial antivirus software effectiveness: An empirical study.
\newblock {\em Computer}, 44(3):63--70, March 2011.

\bibitem{horsepillanalysis}
A.~Tappert and T.~O'Connor.
\newblock The horse pill rootkit vs. forcepoint threat protection for linux.
\newblock
  \url{https://blogs.forcepoint.com/security-labs/horse-pill-rootkit-vs-forcepoint-threat-protection-linux},
  November 2016.
\newblock [Online; accessed: 2018-01-11].

\bibitem{elfspec}
TIS Committee.
\newblock {\em Tool Interface Standard ({TIS}) Executable and Linking Format
  ({ELF}) Specification Version 1.2}, May 1995.
\newblock [Online].

\bibitem{TNC}
T.~TNC.
\newblock Tnc architecture for interoperability version 1.5, revision 3.
\newblock {\em TCG specification}, 1, 2012.

\bibitem{sandbox}
A.~Viswanathan and B.~Neuman.
\newblock A survey of isolation techniques.
\newblock Draft Copy, Information Sciences Institute, University of Southern
  California, 2009.

\bibitem{artofmemoryforensics}
A.~Walters, J.~Levy, A.~Case, and M.~H. Ligh.
\newblock {\em The Art of Memory Forensics: Detecting Malware and Threats in
  Windows, Linux, and Mac Memory}.
\newblock John Wiley \& Sons, Indianapolis, 2014.
\newblock [Online] Available: Safari e-book.

\bibitem{sysjail-breakout}
R.~N. Watson.
\newblock Exploiting concurrency vulnerabilities in system call wrappers.
\newblock In {\em Proceedings of the first USENIX Workshop On Offensive
  Technologies}, pages 2:1--2:8, Berkeley, CA, USA, 2007. USENIX Association.

\bibitem{purelinuxthreadsdemo}
C.~Wellons.
\newblock Pure linux threads demo.
\newblock
  \url{https://github.com/skeeto/pure-linux-threads-demo/blob/master/threads-x86_64.s},
  2015.
\newblock [Online; accessed: 2018-01-29].

\bibitem{evanescentbat}
G.~Wicherski.
\newblock Syscan'14 singapore: Linux memory forensics a real life case study by
  georg wicherski.
\newblock \url{https://www.youtube.com/watch?v=JpY88tnqPhw}, May 2014.
\newblock [Online; accessed: 2018-01-29].

\bibitem{avt}
T.~Wilson.
\newblock Move over, apts -- the ram-based advanced volatile threat is spinning
  up fast.
\newblock
  \url{https://www.darkreading.com/vulnerabilities---threats/move-over-apts----the-ram-based-advanced-volatile-threat-is-spinning-up-fast/d/d-id/1139211?},
  February 2013.
\newblock [Online; accessed: 2018-03-25].

\bibitem{livingofftheland}
C.~Wueest and H.~Anand.
\newblock Internet security threat report: Living off the land and fileless
  attack techniques. an istr special report.
\newblock
  \url{https://www.symantec.com/content/dam/symantec/docs/security-center/white-papers/istr-living-off-the-land-and-fileless-attack-techniques-en.pdf},
  July 2017.
\newblock [Online; accessed: 2018-01-11].

\bibitem{malware-obfuscation}
I.~You and K.~Yim.
\newblock Malware obfuscation techniques: A brief survey.
\newblock In {\em 2010 International Conference on Broadband, Wireless
  Computing, Communication and Applications}, pages 297--300, Nov 2010.

\end{thebibliography}
\bibliographystyle{abbrv}

\end{document}